\documentstyle[prl,aps,twocolumn,psfig,floats]{revtex}
\input{epsf}
\begin{document}
\draft
\twocolumn[\hsize\textwidth\columnwidth\hsize\csname@twocolumnfalse\endcsname

\title{The Superconductor-Insulator Transition in a Tunable Dissipative Environment}
\author{Karl-Heinz Wagenblast$^{a,b}$, Anne van Otterlo$^{b}$, 
	Gerd Sch\"{o}n$^{a}$, and Gergely T. Zim\'{a}nyi$^{b}$}
\address{a) Institut f\"{u}r Theoretische Festk\"{o}rperphysik,
	Universit\"{a}t Karlsruhe, D-76128 Karlsruhe, FRG\\
	b) Physics Department, University of California,
	Davis, CA 95616, USA}
\date{April 1997}
\maketitle

\begin{abstract}
We study the influence of a tunable dissipative environment on the
dynamics of Josephson junction arrays near the superconductor-insulator 
transition.  The experimental realization of the environment is a two 
dimensional electron gas coupled capacitively to the array. This setup 
allows for the well-controlled tuning of the dissipation by changing 
the resistance of the two dimensional electron gas.
The capacitive coupling cuts off the dissipation at low frequencies.
We determine the phase diagram and calculate the temperature and
dissipation dependence of the array conductivity.  We find good 
agreement with recent experimental results.
\end{abstract}
]

Quantum phase transitions attract intense attention because of their
paradigmatic nature: they are relevant to a host of experimental
issues. Examples include the superconductor-insulator transition in
granular superconductors\cite{films}, the transition between Quantum
Hall states\cite{leekivelson}, transitions in disordered
magnets\cite{sachdev}, and the physics of vortices in the presence of
columnar disorder\cite{vinokur}.  Josephson junction arrays constitute
a particularly attractive experimental testing ground for the
superconductor-insulator (SI) transition, because all parameters are
well under control, and are widely tunable\cite{arrays,rimberg97}. In
these systems the SI transition can be driven by quantum fluctuations
when the charging energy $E_C$ becomes comparable to the Josephson
coupling energy $E_J$~\cite{doniach81}. It was understood early that
dissipation is also capable of driving an SI transition.  The phase
diagram of a single Josephson junction in a dissipative environment
was explored by Schmid\cite{schmid83}. Strong dissipation suppresses
quantum fluctuations and restores the classical behaviour with a
finite supercurrent. For weak damping, however, quantum fluctuations
suppress the supercurrent to zero. When an array is built from the
junctions, at strong dissipation phase fluctuations are again damped,
favouring phase coherence and global superconductivity. This type of
SI transition is present in arrays of Josephson junctions as well as
in thin films\cite{chakravarty86,fisher87,emery95,wagenblast97}.

The experimental verification of a dissipation tuned
superconductor-insulator transition is still open.  The actual
strength of the dissipation is hard to control.  An indicator may be
the normal state resistance, although it is unclear how this
translates into a dissipation below the bulk transition temperature,
where the opening of a gap freezes out the gapless
excitations\cite{ambegaokar82}.  It is also unsettled whether the
dissipation or the Coulomb interaction is the main driving force for
the transition.  Recently the Berkeley group succeeded to fabricate
and investigate Josephson junction arrays with tunable dissipation by
placing a Josephson junction array on top of a two dimensional
electron gas (2DEG), separated by an insulator\cite{rimberg97}.  The
electron density and sheet resistance of the 2DEG are varied by tuning
a gate voltage, without influencing the other parameters of the
array. The main result is that the array resistance exhibits a
temperature dependence, parametrized by the dissipation, which is
reminiscent of a superconductor-insulator transition
\cite{films,arrays}.

In the present work we model the experimental setup by an array
capacitively coupled to a 2DEG. Our results for the array resistance
per square $R(T)$ as a function of temperature $T$ and 2DEG resistance
$R_{\text{2DEG}}$ track the experimental data well.  The $T$
dependence of $R(T)$ is characteristic of an imminent SI transition tuned by
$R_{\text{2DEG}}$. However, at the lowest temperatures it exhibits a
sharp reentrant rise. In our model this is naturally explained by the
presence of a cutoff in the spectrum of the dissipation at low
frequencies.

The quantum-dynamical variables describing a Josephson junction array
are the phases $\varphi_j$ of the superconducting order parameter
on island $j$. 
The dynamics of the 2DEG is formulated in terms of a fluctuating scalar
potential $V({\bf r},t)$~\cite{ambegaokar82}.  
This potential can, in close analogy to the Josephson relation, be
represented by a phase-variable $\phi({\bf r},t)$, 
defined by $\hbar\dot{\phi}({\bf r},t)=2eV({\bf r},t)$.
For the coupled system the action takes the form 
\begin{eqnarray}
	S[\varphi,\phi]=S_{\text{JJA}}[\varphi]
	+S_{\text{I}}[\varphi,\phi]+S_{\text{2DEG}}[\phi]\,.
\end{eqnarray}
The array is characterized by the Josephson coupling $E_{J}$
and the inter-grain capacitance $C_{1}$, which represents the bare, unscreened
Coulomb interaction in 2 dimensions
\begin{eqnarray}
	\nonumber
	S_{\text{JJA}}[\varphi]&=&\frac{1}{2}\int_{k,\omega_\mu}
	\frac{C_{1}}{4e^2}\gamma(k)\omega_\mu^2|\varphi_{k,\omega_\mu}|^2\\
	&&-E_{J}\int d\tau\sum_{<ij>}\cos(\varphi_{i}-\varphi_{j})\,,
\end{eqnarray}
where $\gamma(k)$ is, on a square 2d lattice, given by 
$\gamma(k)=4-2\cos(ak_x)-2\cos(ak_y)$, $a$ is the lattice constant, chosen 
as the unit of length.  At the relevant long wavelengths 
$\gamma(k)\approx k^2$.  
Coupling to the 2DEG is characterized by a capacitance $C_{0}$
\begin{eqnarray}
	S_{\text{I}}[\varphi,\phi]=\frac{1}{2}\int_{k,\omega_\mu}
\frac{C_0\omega_\mu^2}{4e^2}|\varphi_{k,\omega_\mu}-\phi_{k,\omega_\mu}|^2\, ,
\end{eqnarray}
where $\dot{\varphi}-\dot{\phi}$ is the potential {\it difference}
between the array and the 2DEG.  This formulation of an interaction
mediated by a local capacitance correctly represents the
electrostatics of the system.  The interaction between the charges in
the array and the 2DEG is determined by the {\em inverse} capacitance
matrix which represents a long-range interaction of charges.

The dynamics of the 2DEG is ohmic, with resistance $R_{\text{2DEG}}$.  
The microscopic details of the 2DEG do not play a role on the length scales
considered presently. The corresponding action is
\begin{equation}
	S_{\text{2DEG}}[\phi]=\frac{1}{2}\int_{k,\omega_\mu}
	\frac{R_Q}{2\pi R_{\text{2DEG}}}k^2|\omega_\mu|
	|\phi_{k,\omega_\mu}|^2\,.
\end{equation}
The scale of resistance is set by its natural quantum unit, $R_Q=h/(4e^2)$.
The interactions within the 2DEG, and its diffusive behaviour
influence the action only at higher momenta and frequencies.

The effective action for the array is constructed by integrating
out $\phi$, the fluctuating voltage of the 2DEG
\begin{eqnarray}
	\nonumber
	S_{\text{eff}}[\varphi]&=&\frac{1}{2}\int_{k,\omega_\mu}D^{-1}_{0}
	(k,\omega_\mu)|\varphi_{k,\omega_\mu}|^2\\
	&&-E_J \int d\tau \sum_{<ij>}\cos(\varphi_i-\varphi_j) ~~.
\end{eqnarray}
The propagator for the $\varphi$ reads
\begin{eqnarray}
	\label{prop}
	D^{-1}_{0}(k,\omega_\mu)=\frac{C_1}{4e^2} k^2\omega_\mu^2 
	+\frac{C_0}{4e^2}\frac{k^2\omega_{\mu}^2}{k^2+|\omega_\mu|/\Omega_0} ~~,
\end{eqnarray}
where $1/\Omega_0=R_{\text{2DEG}} C_0$.  The dynamics of the phases
$\varphi$, described by the propagator $D_0$, has three characteristic
frequency regimes:\\
$1)$ In the limit of small frequencies,
$\omega < \Omega_0$,  dissipation is frozen out, and 
the dynamics of the phase is capacitive. In this limit 
Eq.(6) reduces to $D_0^{-1}=\omega_\mu^2(C_1 k^2 + C_0)/(4e^2)$.  
The 2DEG screens the electrostatic
interaction in the array beyond the characteristic length scale
$\Lambda=\sqrt{C_1/C_0}$.  \\
$2)$ At frequencies exceeding $\Omega_0$, the resistivity of the 2DEG
induces damping for the superconducting phase.  The origin of this
damping is that the voltage fluctuations of the 2DEG cannot follow the
fluctuations of $\varphi$ adiabatically. This creates damping with a
strength determined by the 2DEG resistance.  From Eq.(6) one finds
$D_0^{-1}=C_1 k^2\omega_\mu^2/(4e^2)+k^2|\omega_{\mu}|R_Q/(2\pi
R_{\text{2DEG}})$, (for $\omega > \Omega_0$), which describes an array
of resistively shunted junctions\cite{fisher87}.\\
$3)$ At even higher frequencies
$\omega_{\mu}\gg\Omega_1=1/(R_{\text{2DEG}} C_1)$, the response is
again capacitive, but now determined by the inter-grain capacitance
$C_1$.  The leading frequency dependence of the propagator of Eq.(6)
is $D_0^{-1}=\omega_\mu^2 k^2 C_1/(4e^2)$.

In sum, the effective action for the array is ohmic only in an {\it
intermediate} frequency range $\Omega_0 <\omega_{\mu}< \Omega_1$.  At
the lowest and highest frequencies the dynamics is capacitive.  The
two energy scales are well separated in the case $C_0\gg C_1$.  A
quantum phase transition is driven by the behaviour of the action at
the lowest frequencies.  In the present case - as the dissipative
action is cut off at the lowest frequencies - a dissipation driven
transition cannot occur in the strict sense.  However, a
quasi-critical behaviour can be observed at temperatures and voltages
exceeding the low energy scale $\Omega_0$.  In the limit
$\Omega_0\rightarrow 0$ ($C_0\rightarrow\infty$) this behaviour
converges to the true dissipation-tuned transition.

To characterize this quasi-critical behaviour, we now evaluate the
electromagnetic response of the array at finite temperatures.  The
array conductivity, as a function of Matsubara frequencies, is
calculated via the Kubo formula.  In the regime where the Josephson
energy $E_{J}$ is smaller than the capacitive energy scale
$E_{C}=e^2/(2C_0)$, insight can be gained by a perturbative expansion to
second order in $E_J$.  For the longitudinal part of the conductivity
we obtain
\begin{eqnarray}
	\label{mats}
  	\sigma_{xx}(\omega_\nu) = \frac{2e^2 E_J^2}{\hbar}\int_0^{\beta} 
  	d\tau \frac{1-\mbox{e}^{i\omega_\nu \tau}}{\omega_\nu} g(\tau)\,,
\end{eqnarray}
where $g(\tau)=\langle\cos[\varphi(0,\tau)
-\varphi(\hat{x},\tau)-\varphi(0)+\varphi(\hat{x},0)]\rangle_0$.
The correlation function $g(\tau)$ depends on $\hat{x}$, which is defined to
connect the nearest neighbors in the $x$ direction. 
The expectation value $\langle ...\rangle_{0}$ is
taken with the action $S_{\text{eff}}[\varphi]$ at $E_J=0$.  
The result is 
\begin{eqnarray}
	\nonumber
	g(\tau)&=&\exp\left\{\frac{1}{\beta}\sum_{\mu}\,(1-\cos\omega_\mu\tau)
	\,d(\omega_\mu)\right\}\cdot g_W(\tau)\\ \nonumber
	d(\omega_\mu)&&=\int \frac{d^2k}{4\pi^2}(2-2\cos k_x)D_0(k,\omega_\mu)\\
	&&\approx\frac{2\pi}{\alpha|\omega_\mu|}\frac{1}{1+|\omega_\mu|
	/\Omega_1} + \frac{E_0}{\omega_\mu^2}\, ~,
\end{eqnarray}
where $\alpha = R_Q/R_{\text{2DEG}}$,
$E_0=2\pi\Omega_0/\alpha$, and $g_W(\tau)$ represents a
summation over the winding numbers, which reflects the discrete nature
of the charge transfer in the array\cite{halperin}.  In accordance
with the above, $d(\omega_{\mu})$ represents ohmic damping in the
intermediate frequency range $\Omega_{0}<\omega_{\mu}<\Omega_{1}$.
The lattice structure and the range of the electrostatic interaction
influence the precise value of the damping strength $\alpha$
upto a multiplicative constant $c$. For a square array and short
range interactions ($C_0\gg C_1$) this prefactor is $c\sim{\cal O}(1)$.

The conductivity as a function of real frequencies
follows by analytic continuation, $\sigma(\omega)=\sigma_{xx}(
\omega_\mu\rightarrow -i\omega+\delta)$
\begin{equation}
	\sigma(\omega)=\frac{2\pi E_J^2}{R_Q}\int_0^{\infty} 
        dt \frac{1-\mbox{e}^{i\omega t}}{-i\omega}\mbox{Im}[g(it)]
\end{equation}
The analytic continuation of $g(\tau)$ reads
\begin{eqnarray}
	\nonumber
	g(it)&=&\exp\bigg(-\frac{2}{\alpha}\int_0^{\infty} d\omega\Big[
	\frac{1-\cos\omega t}{\omega(1+\omega^2/\Omega_1^2)}
	\coth\frac{\beta\omega}{2}\\ && -i\frac{\sin\omega t}
	{\omega(1+\omega^2/\Omega_1^2)}\Big]-iE_0t\bigg)\cdot
         \langle \cos (ntE_0)\rangle_{n}\,.
\end{eqnarray}
Here $\langle ...\rangle_{n}$ is taken with the action $S_n=
{\beta}E_0 n^2/2$, $n$ integer.  We identify the energy scale $E_0$
with the Mott gap, the energy cost of adding on extra charge to the
array.  The Mott gap $E_0$ is considerably reduced by random offset charges 
and, close to the SI transition, also by charge fluctuations\cite{otterlo93}.  
For low temperatures $T \ll \Omega_1$ the correlator is given by
\begin{eqnarray}
	\nonumber
	\mbox{Im}\,g(it)&=&\left(\frac{1-\mbox{e}^{-2\pi t/\beta}}
	{2\pi t/\beta}\right)^{-\frac{2}{\alpha}}
	\sin\left(\frac{\pi}{\alpha}(1-\mbox{e}^{-\Omega_1 t})+E_0t\right)\\
	\nonumber
        &&\exp\bigg[-\frac{1}{\alpha}\bigg(
	2\gamma +2\ln(\Omega_1 t)+\mbox{e}^{\Omega_1 t}
	\mbox{E}_1(\Omega_1 t)\\
	&& -\mbox{e}^{-\Omega_1 t}\mbox{Ei}(\Omega_1 t)\bigg)
	-\frac{2\pi t}{\alpha\beta}\bigg] \langle \cos (ntE_0)\rangle_{n}\,,
\end{eqnarray}
where $\gamma=0.577...$ is Euler's constant, Ei and E$_1$ are
exponential integrals.  A subsequent numerical integration directly
gives the conductivity as shown in Fig.1, for various values of
$R_{\text{2DEG}}$.  The conductivity $\sigma(T)$ in the {\em
intermediate} temperature range $\Omega_{0}<T<\Omega_1$ behaves as
$\sim T^{2/\alpha-2}$, in analogy to the single junction
result\cite{schmid83,fisher87}. For $\alpha > 1$ the decrease of
$R(T)$ indicates an impending SI transition. However, since the
dissipation is frozen out below $\Omega_{0}$, $R(T)$ rises sharply,
eventually becoming an insulator at $T=0$. For $\alpha < 1$, $R(T)$ is
monotonously increasing with decreasing $T$.
\begin{figure}[hbt]
  \unitlength1cm 
  \begin{picture}(8.3,6.5)
  \put(0,0.2){\psfig{figure=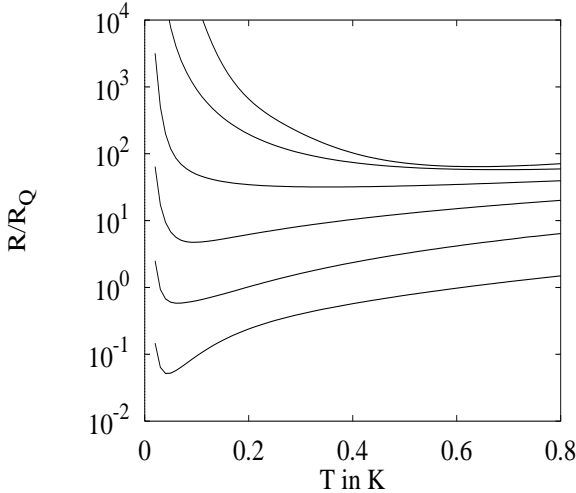,height=6.5cm,width=8.0cm}}
  \end{picture} 
  \caption{Temperature dependence of the array
  resistivity $1/\sigma(\omega$=0). $E_0$=0.2\,K, $E_J$=0.28\,K,
  $C_0/C_1$=10. The dissipation takes the values $\alpha$=20 (lowest
  curve), 5, 2, 1, 0.5, 0.25 (uppermost curve).}
\end{figure}

These perturbative results can be interpreted as a renormalization of
$E_J$ by the dissipative processes. Writing $\sigma(T) \sim
(E_{J}^{\text{ren}})^2 $ identifies $E_{J}^{\text{ren}} \sim
T^{1/\alpha-1}$. This renormalization stops at $\sim \Omega_{0}$. At
$\Omega_{0}$ the model is equivalent to an $XY$ model, with
renormalized parameters. This model has a phase transition at
$E_{J}^{\text{ren}}/E_{C} \approx 1$. Therefore, the SI phase boundary
is given by $E_{J}/E_{C} \approx (C_{1}/C_{0})^{1-1/\alpha}$.  With
decreasing $\Omega_0\rightarrow 0$ $(C_0\rightarrow\infty)$ a true
dissipation driven phase transition is approached at $\alpha=1$.  For
small $\alpha$, the dissipative scaling is taken over by the XY
scaling\cite{wagenblast97}. The phase boundary flattens and
reaches smoothly its $\alpha=0$ value of ${\cal O}(1)$ (Fig.2).  For
large $E_{J}$ we recall the results of Ref.\cite{chakravarty86}, which
established that in this limit the effective action reduces to that of
a single junction. Consequently, the renormalization group flows,
obtained perturbatively at small $E_{J}$ are characterized by the same
power laws at large $E_J$.  Thus in this regime for large $\alpha$ the
resistivity decreases monotonically and the array becomes truly
superconducting.  For small $\alpha$ the RG flows are not completely
clear. In the very large $E_J$ limit one expects the dominance of
single junction physics at intermediate scales, accompanied with a
rise of the resistance. At lower $T$ collective processes drive the
system superconducting, manifesting themselves in a sharp drop of the
resistivity. The phase diagram and the temperature dependences of the 
resistance in the four regimes are summarized in Fig.2. 
\begin{figure}[hbt]
  \unitlength1cm
  \begin{picture}(8.3,6.5)
   \put(0,0.2){\psfig{figure=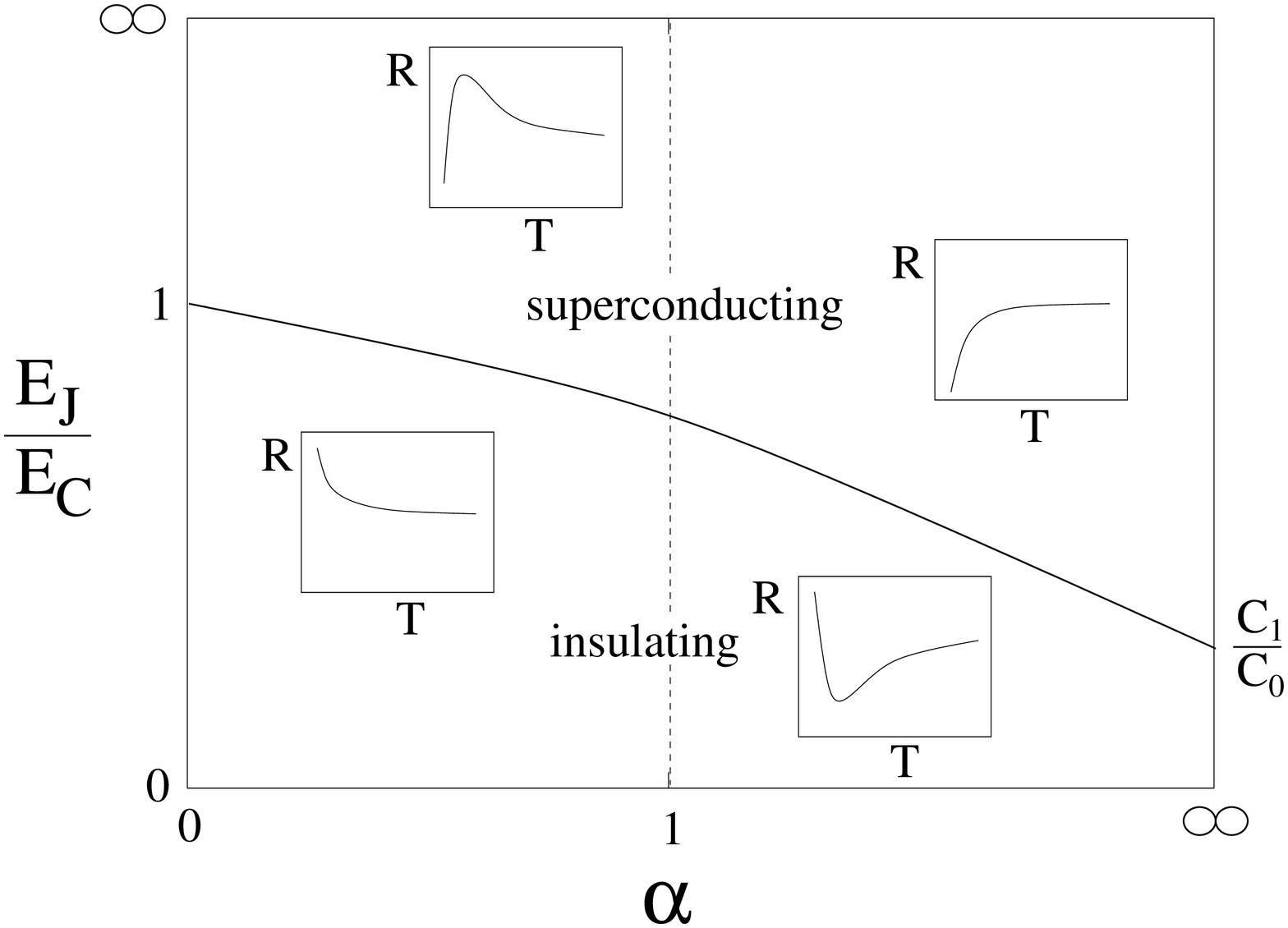,height=6.5cm,width=8.0cm}}
  \end{picture}
  \caption{Phase diagram of an array coupled capacitively to a 2DEG. 
The insets show $R(T)$ as a function of the temperature in the
different regions.}
\end{figure}

Our perturbative analysis concentrated on the low temperature
behaviour of the resistivity. It does not include quasiparticle
currents.  The normal state resistance $R_N$ at higher temperatures
can be only reproduced by including {\it thermally activated}
quasiparticles.  They form a parallel channel to the flow of the
Cooper pairs.  Using the standard BCS gap $\Delta(T)$ introduces
visible change only close to the bulk $T_{c0}$. However, it was
recently argued that phase-space considerations seriously {\it reduce}
the gap, experienced by the quasiparticles \cite{feigelman}. 
In Fig.3, $R(T)$ is plotted with such a parallel normal channel, 
using a reduced gap value of $E_g=0.2$K. $R(T)$ now exhibits
a convergence to the normal state resistivity $R_N$ at higher temperatures.

\begin{figure}[hbt]
  \unitlength1cm
  \begin{picture}(8.3,6.5)
   \put(0,0.2){\psfig{figure=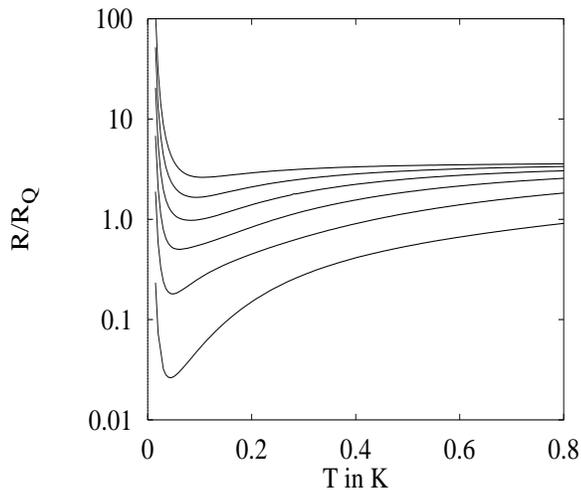,height=6.5cm,width=8.0cm}}
  \end{picture}
  \caption{Array resistivity $1/\sigma(\omega$=0) with a parallel
thermally activated channel $R_N\exp(E_g/T)$, with $E_g$=0.2\,K.
$E_0$=0.2\,K, $E_J$=0.28\,K, $C_0/C_1$=10, and $R_N$=23\,k$\Omega$.  
At $R_{\text{2DEG}}$= 200, 700, 1200, 1700, 2200, 2700\,$\Omega$.}
\end{figure}

Recent experiments carefully explored Josephson junction arrays
capacitively coupled to a 2DEG\cite{rimberg97}. The experiments fall
in the parameter regime, where $E_J/E_C$ is small and our perturbative
analysis is applicable.  The temperature dependence of the resistivity
$R(T)$ is strikingly similar to that in Fig.3. Also, the dependence of
the array resistance on $R_{\text{2DEG}}$ at fixed temperatures was
determined.  Since $R_Q/R_{\text{2DEG}} \sim \alpha$, the power law
dependence $R(T) \sim T^{2-2/\alpha}$ translates into an exponential
relation between $R$ and $R_{\text{2DEG}}$ (Fig.4).  This is again in
good agreement with the experiments\cite{rimberg97}.
\begin{figure}[hbt]
  \unitlength1cm
  \begin{picture}(8.3,6.5)
   \put(0,0.2){\psfig{figure=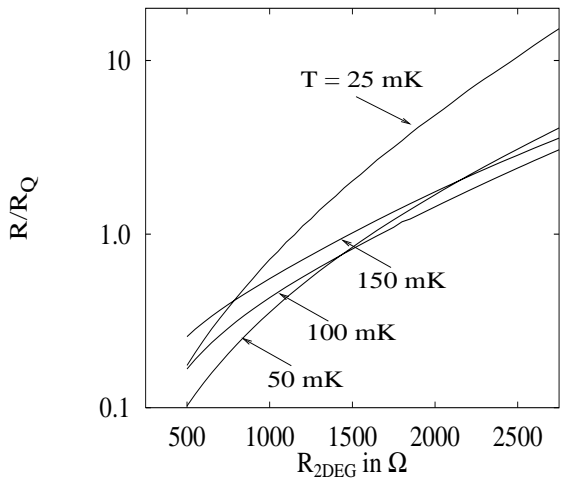,height=6.5cm,width=8.0cm}}
  \end{picture}
  \caption{Array resistivity $1/\sigma(\omega$=0) as a function of 
$R_{\text{2DEG}}$. $E_0$=0.2\,K, $E_J$=0.28\,K, $C_0/C_1$=10, 
at $T$=25, 50, 100, 150\,mK.}
\end{figure}

Lastly, let us consider an Ohmic coupling between the array and the
2DEG.  In this case the effective action takes the form of the
resistively shunted junction model. The resistances of the 2DEG and the
Ohmic shunt between an island and the 2DEG are in series. Since now
the spectrum is Ohmic down to zero frequency, the dissipation drives a
{\it true} SI transition. This may be realized by doping the
semiconducting layer which separates the array from the 2DEG. In this
arrangement a local damping of the phase is generated via the Andreev
process, which allows Cooper pairs to decay into normal electrons
in the substrate\cite{wagenblast97,geshkenbein97}.

In summary, we developed a model for a Josephson junction array
capacitively coupled to a two dimensional electron gas. We determined
the phase diagram and the corresponding dependence of the resistivity
on the temperature and the resistivity of the 2DEG. Our results
compare well to the recent experimental data of Ref.\cite{rimberg97}.
We also suggested further experiments to investigate the
dissipation-tuned phase transition more closely.

We thank J. Clarke, R. Fazio, T.R. Ho, A. Huck, C. Kurdak and A.J. Rimberg for 
valuable discussions. This research was supported by NSF DMR 95-28535, 
by the SFB 195 of the DFG, and by the DAAD.

\end{document}